\documentclass[a4paper,11pt]{article}
\pdfoutput=1 % if your are submitting a pdflatex (i.e. if you have
\usepackage{jinstpub} % for details on the use of the package, please
\usepackage{url}
\usepackage{float}
\usepackage{orcidlink}

\title{\boldmath Nuclear fuel imaging using position-sensitive detectors}

\author[1]{S. Saariokari\orcidlink{0000-0002-6798-2454}\note{Corresponding author.}}
\author{, P. Dendooven\orcidlink{0000-0003-1859-1315}}
\author[2]{, M. Laassiri\orcidlink{0000-0001-7146-4468}\note{Present address: Department of Physics, Brookhaven National Laboratory, Upton, NY 11983, USA.}}
\author{and J. E. Brücken\orcidlink{0000-0001-6066-8756}}

\affiliation{Helsinki Institute of Physics, University of Helsinki, Finland}

\emailAdd{santeri.saariokari@helsinki.fi}

\abstract{
We are evaluating the performance of a Passive Gamma Emission Tomography (PGET) device \cite{iaea} equipped with 3D position-sensitive cadmium zinc telluride (CZT) gamma-ray detectors when used for inspecting spent nuclear fuel assemblies (SFAs). Before their disposal in a geological repository, SFAs undergo verification using the PGET device, which was developed under the guidance of the IAEA and approved by the IAEA for safeguards inspections. Recent advancements in imaging detector technology may offer a method to extend the capabilities of such devices beyond standard safeguards applications, allowing an efficient non-invasive way to accurately characterise the properties of nuclear fuel assemblies.
The efficiency of the currently used small CZT detectors is restricted by the limited likelihood of full gamma-ray absorption, which is needed for optimal imaging information. Employing larger CZT detectors would increase the probability of capturing the full energy of gamma rays, thereby enhancing the sensitivity of the PGET device and the quality of the reconstructed images. Large CZT detectors need to be position-sensitive to determine through which collimator slit a gamma ray travelled. Position sensitivity results from the pixelated readout of the CZT crystals. Pixelation potentially increases the spatial resolution of the system, which is currently determined by the collimator used. Pixelation allows resolving the position of arrival up to (readout pitch)/$\sqrt{12}$. We are additionally exploring the potential of utilising Compton imaging to provide information on the origin of gamma rays along the SFA.
Monte Carlo simulations are used to estimate the increase in full photon absorption efficiency when comparing large and current, small, crystals. A dedicated simulation is created using Geant4, where gamma rays of energy 661.7 keV and 1274 keV are targeted to the model describing the approved apparatus now equipped with 22 mm x 22 mm x 10 mm crystals of CZT. It is observed that the efficiency for photon absorption in this case is greatly increased when compared to the existing detectors.
}


\proceeding{25$^{\text{th}}$ International Workshop on Radiation Imaging Detectors\\
  30 June - 4 July\\
  Lisbon, Portugal}

\begin{document}
\maketitle
\flushbottom

\section{Introduction}
\label{sec:intro}

Finland will start storing its spent nuclear fuel in a deep geological repository in 2025, with Passive Gamma Emission Tomography (PGET) \cite{iaea} measurements being an integral part of nuclear safeguards at the ONKALO repository \cite{posiva}. The International Atomic Energy Agency (IAEA) approved the design of a PGET device in 2017 to characterize spent nuclear fuel assemblies (SFAs) for nuclear safeguards purposes. The design has been tested extensively in recent years, showing capability in identifying fuel rods missing from the assembly \cite{virta2020,virta2023}.  This study aims to improve the PGET device's design, improving the performance of nuclear safeguards as implemented by the IAEA. Recent advancements in detector technology may offer a method to extend the capabilities of said device, also beyond standard safeguard applications. Employing larger cadmium zinc telluride (CZT) detectors increases the probability of capturing the incoming gamma rays as well as secondary particles produced in the interaction with the crystal, enhancing the sensitivity of the PGET device and the quality of the reconstructed images. Pixelation improves the system's spatial resolution, possibly enabling Compton imaging.

The quantity typically used for tomographic imaging of spent nuclear fuel is the intensity of the $^{137}$Cs gamma peak at 661.7~keV energy. Another possible quantity for this is the $^{154}$Eu gamma peak at 1274~keV. Although far less prominent, the higher energy gamma photons are attenuated significantly less by the fuel rods containing uranium oxide. Attenuation caused by the fuel rods is one of the current challenges of imaging fuel assemblies \cite{virtacentral}. Due to their lower detection probability combined with a lower occurrence, $^{154}$Eu gamma rays cannot be utilised in gamma tomography with current methods.

PGET imaging of spent nuclear fuel is based on a parallel slit collimator gamma camera, selecting the travel direction of gamma rays seen by the detectors, thereby providing imaging information. The spatial resolution of the IAEA device is determined by its collimator pitch, as the detection system consists of one single-readout CZT crystal per collimator slit. Having the identical approach with the larger crystals is impossible due to the crystal size being much larger than any practical collimator slit pitch. The solution is to use crystals with a pixelated readout, which allows position-sensitive readout. The commercial detector GDS-100 manufactured by IDEAS (Oslo, Norway) is used as a reference device, supporting large CZT sensors with a readout pixel pitch of 2~mm in an $11\times11$ grid \cite{gds-100-compton}. It also features the possibility to read up to nine-pixel clusters centred around a triggered anode, mitigating the issue of charge sharing.

Simulations are used to estimate the increase in full photon absorption efficiency when comparing large and current, small, crystals. A dedicated simulation is created using Geant4, where gamma rays of energy 661.7 keV and 1274~keV are targeted to the model describing the approved apparatus now equipped with 22 mm x 22 mm x 10 mm crystals of CZT. Two different configurations of large crystals are considered, each with its own advantages. The viability of using the large crystals for Compton imaging is evaluated in two steps, first ignoring the limits in position and energy resolution, and then utilizing parameters achievable with the GDS-100.

\section{Simulation description}
\label{sec:simulation}

\begin{figure}[h]
\centering
    \includegraphics[width=0.49\textwidth]{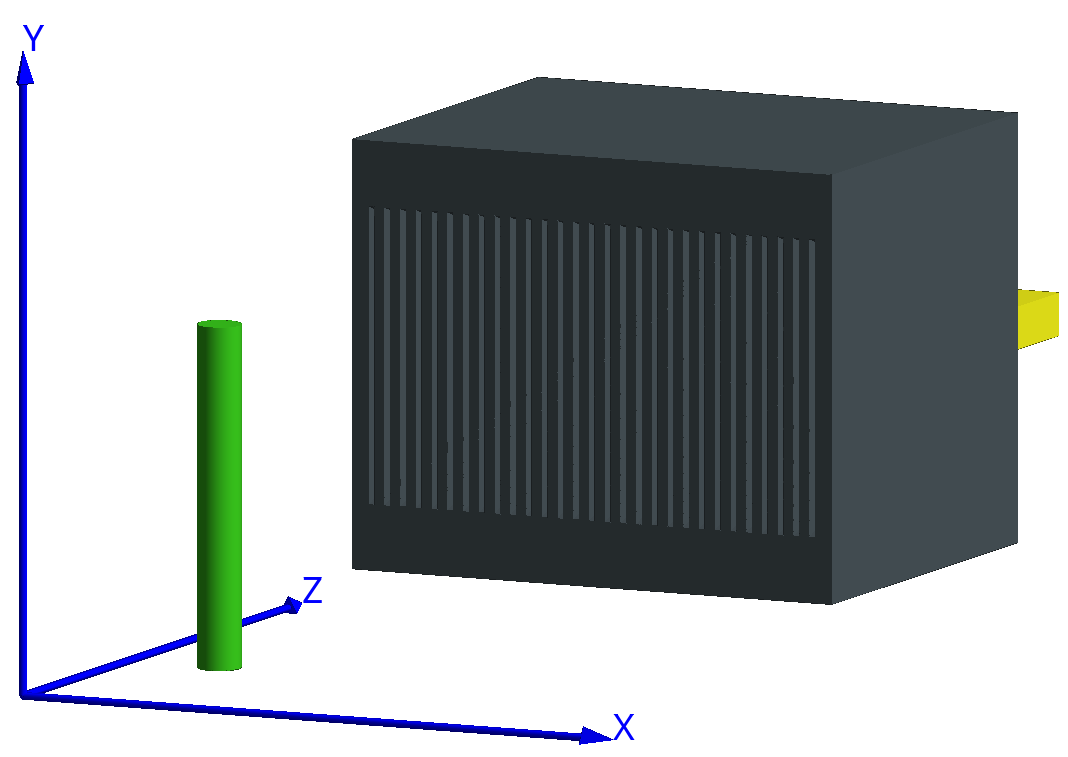}
    \includegraphics[width=0.49\textwidth]{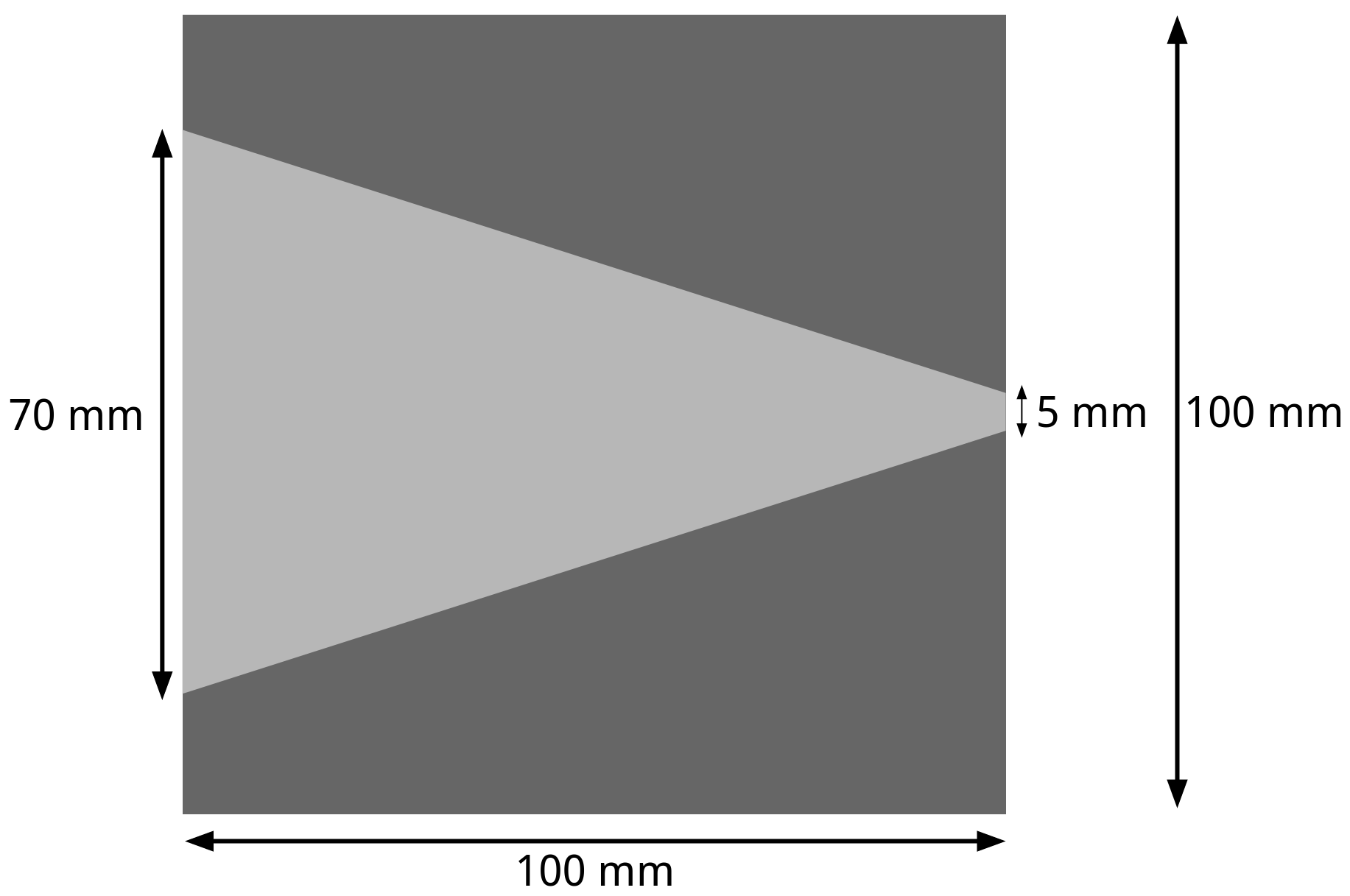}
\caption{Left: Front view of the system, with the source in green in the foreground, the collimator in dark gray and the CZT crystals in yellow. Right: Schematic image of the collimator with the light gray indicating the slit.\label{fig:geometry}}
\end{figure}

A dedicated Monte Carlo simulation is created to compare the detection efficiency of the current and larger CZT crystals for 661.7~keV and 1274~keV gamma rays. The simulation is implemented using the Geant4 framework \cite{geant4}. The geometry consists of a cylindrical monoenergetic gamma emitter, a collimator made of tungsten alloy and a set of CZT crystals, with air as the medium. Attenuation in the source is not taken into account as the target of comparison is the sensitivity to the non-attenuated photons, which yield the best imaging performance. The orientation of the components, and the geometrical details of the collimator are shown in figure \ref{fig:geometry}. The height of the collimator is chosen to be large enough to cover the whole source. The collimator is modelled in detail after the collimator in the PGET device but extends over 10~cm instead of 36~cm in the x-direction. The height of the slits was chosen to match exactly the dimensions of the current device.

\begin{figure}[h]
\centering
    \raisebox{0.7em}{\includegraphics[width=0.28\textwidth]{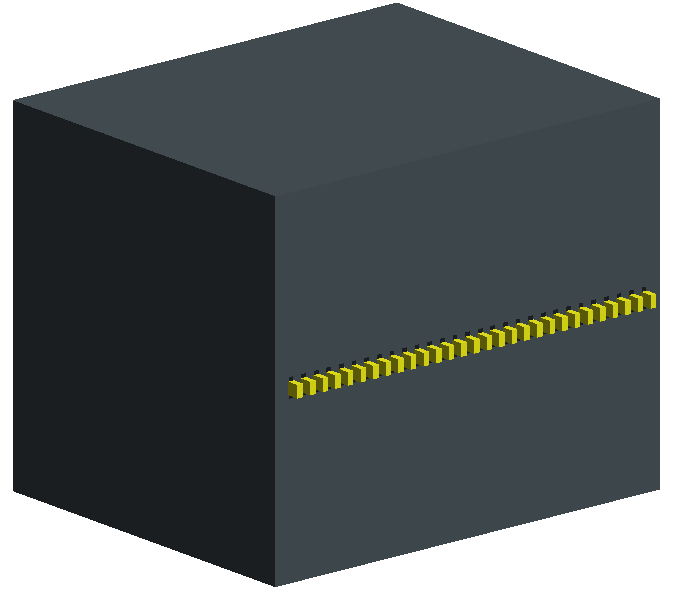}}
    \raisebox{1em}{\includegraphics[width=0.28\textwidth]{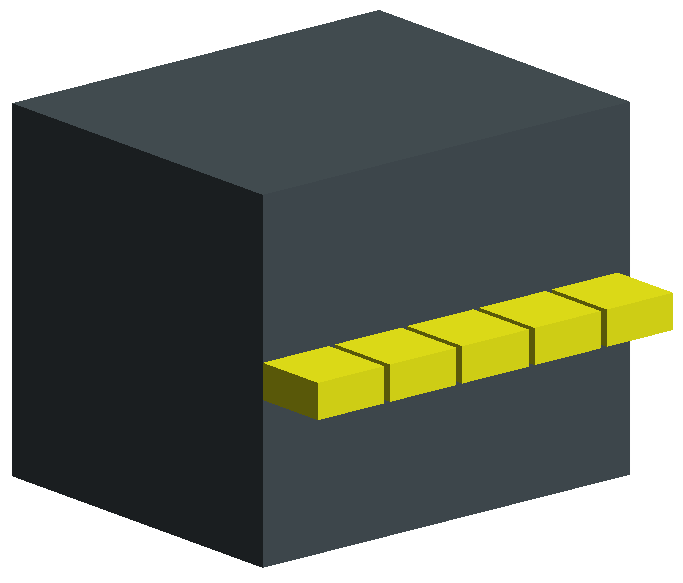}}
    \includegraphics[width=0.40\textwidth]{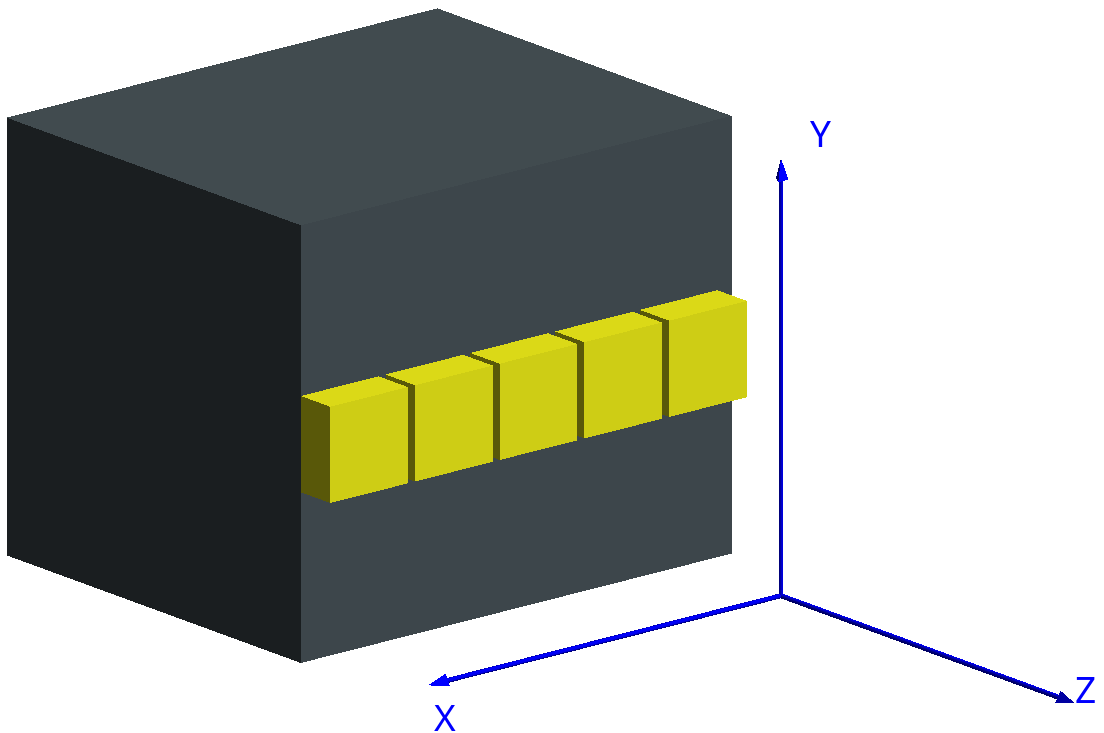}
\caption{Left: current crystal configuration. Center: Large crystals in configuration 1. Right: Large crystals in configuration 2.\label{fig:crystals}}
\end{figure}

\noindent Three different CZT crystal dimensions are considered: the ones featured in the IAEA-approved design \cite{iaea}, and two configurations having the largest crystals compatible with the reference device. The sizes of the crystals are 1.75 mm $\times$ 3.5 mm $\times$ 3.5 mm (original), 22 mm $\times$ 10 mm $\times$ 22 mm and 22 mm $\times$ 22 mm $\times$ 10 mm for the current device, proposed configuration 1 and proposed configuration 2, respectively. These will be referred to as "configuration 1" and "configuration 2" in the future for simplicity. The variations used are depicted in figure \ref{fig:crystals}. It can be noted, that the large crystals extend beyond the collimator slits in the y-dimension. This is desirable, as the secondary particles (scattered photon and electron) from Compton scattering inside the crystal emerge at an angle with respect to the direction of the incident gamma ray. Extending the detector volume laterally increases the likelihood of capturing these particles, improving the spectral performance.

\noindent The collimator material is chosen to be a readily available tungsten alloy TRIAMET~S18~\cite{triamets18}, having mass fractions of 95~\% tungsten, 3.5~\% nickel and 1.5\% copper. The slit pitch is 4~mm with a slit width of 1.5~mm, which allows full collimator coverage with five 22~mm wide crystals. 

A total of 200 million gamma photons are created for each simulation run. A simulation event is defined as all interactions resulting from one primary gamma ray. The resulting photon beam is limited to the general direction of the collimator to improve the simulation performance. The number of primaries is chosen based on reaching convergence rather than mimicking a specific measuring time, prioritizing the comparison of the three detector configurations. The physics libraries enabled in Geant4 are \texttt{G4EmStandardPhysics\_option4} and \texttt{G4DecayPhysics}. The lower limits for creating new secondary particles or "production cuts" are adjusted to accommodate the most common characteristic x-rays of the collimator constituents. The numerical values of the energy cuts are given in table \ref{tab:cuts}.

\begin{table}[h]
    \centering
    \caption{\label{tab:cuts}Secondary particle production cuts in the simulation.}
    \smallskip
    \begin{tabular}{|l|c|c|c|}
         \hline
         Material & $\lambda$ (µm) & $E_\gamma$ (keV) & $E_{e^-}$ (keV) \\
         \hline
         CZT & 10 &  2.4 & 43 \\
         \hline
         Collimator & 10 & 7.2 & 79 \\
         \hline
         Air & 10 & 0.99 & 0.99 \\
         \hline
    \end{tabular}
\end{table}

\noindent The production cuts set a limit for accurate description of energy deposition, as particles with energy lower than the production cut are assumed to stay in the volume they are created in.

\section{Efficiency}
\label{sec:efficiency}

\begin{figure}[h]
    \centering
    \includegraphics[width=\linewidth]{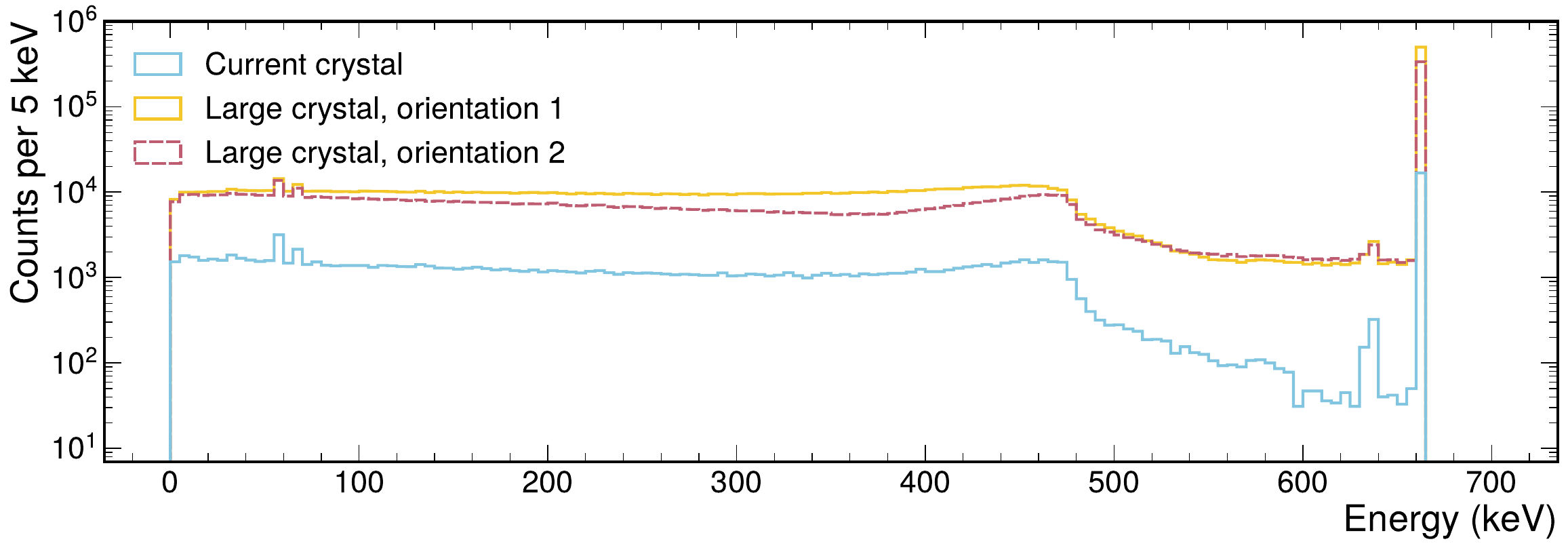}
    \includegraphics[width=\linewidth]{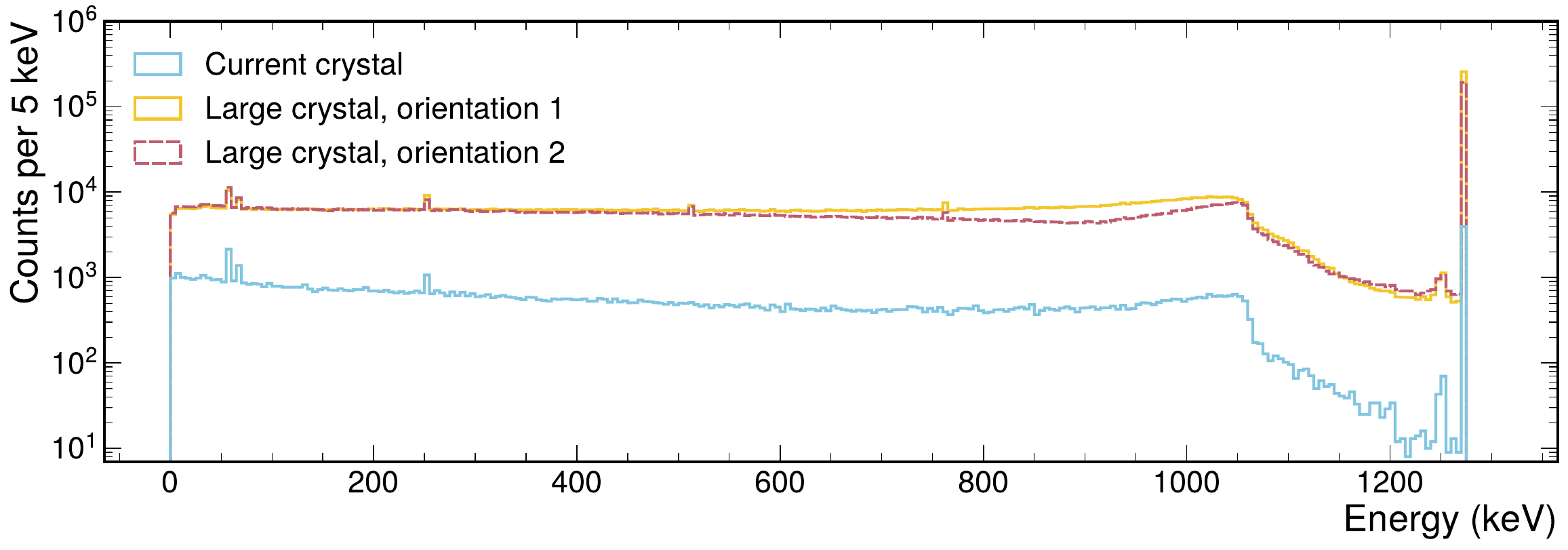}
    \caption{\label{fig:energy}Histogram of the deposited energy per event for $^{137}$Cs (top) and $^{154}$Eu (bottom).}
\end{figure}

The energy spectra obtained from the simulation are studied to examine the differences between the spectroscopic performances of the studied configurations. Figure \ref{fig:energy} shows the histogram of the deposited energy per event for primary photon energies of 661.7~keV ($^{137}$Cs) and 1274~keV ($^{154}$Eu). The expected spectral features include the photo peak, Compton edge and characteristic x-rays from the tungsten collimator. The energy of the Compton edge can be calculated with the simple expression \mbox{\large$E_{e} = \frac{2E_\gamma^2}{2E_\gamma + m_e}$}\vspace{0.2em}, where $E_\gamma$ is the photon energy and $m_e$ the electron mass. The expected location for the Compton edge is 477.4~keV for $E_\gamma=661.7$~keV, and 1061~keV for $E_\gamma=1274$~keV. Tungsten emits characteristic x-rays at 58~keV, 59~keV and 67 keV. Additionally, the annihilation peak at 511~keV is expected to be present in the case of $^{154}$Eu.

\noindent The expected features can be seen in the simulated spectra. Importantly, the height of the photo peak is increased substantially when comparing small and large crystal sizes. The height of the 511~keV annihilation peak seen in the $^{154}$Eu histogram is noted to be small. The $^{137}$Cs peak-to-Compton ratio, or the ratio of the photo peak to the Compton edge, is noted to be 10.4 for the current crystals, 41.4 for configuration 1 and 35.9 for configuration 2. For $^{154}$Eu, the corresponding ratios are 6.2, 29.3 and 25.5.

\begin{figure}[h]
    \centering
    \includegraphics[width=\linewidth]{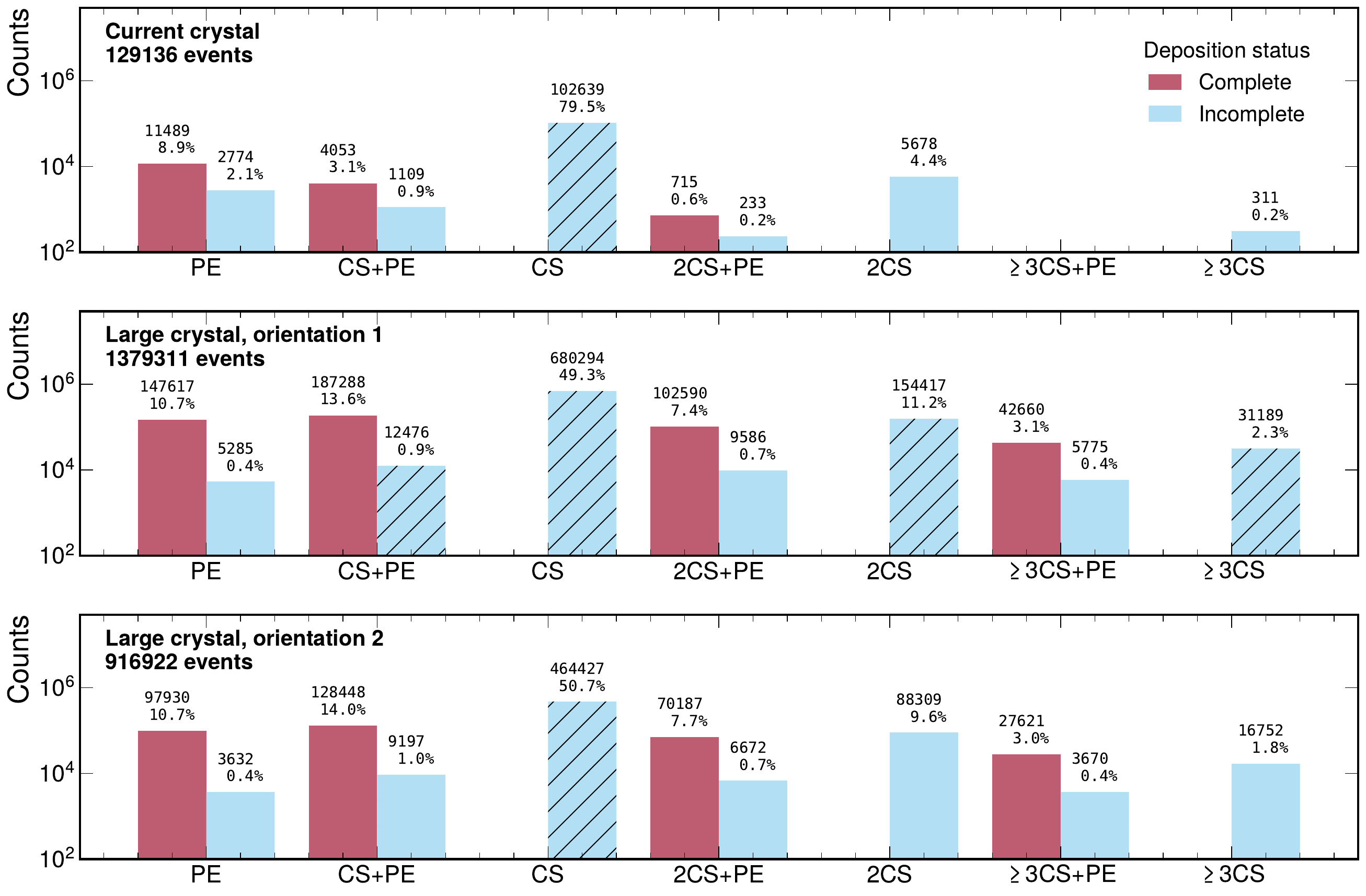}
    \caption{Process frequency in CZT crystals for 661.7~keV primary gamma energy. The number and percentage reading above the bar charts show the absolute and relative frequency of the event category.}
    \label{fig:frequency-cs}
\end{figure}

The viability of Compton imaging depends on the frequency of events where a photon undergoes a Compton scattering and is then absorbed by photoelectric effect. This is referred to as a \textit{Golden event}. Events where multiple Compton scatterings take place can also be utilized, but it is technically more challenging. The efficiency of the three crystal configurations is compared by studying the frequencies of relevant processes. For this, the events are classified according to seven categories: the primary photon is absorbed directly by photoelectric effect (PE), the photon is absorbed after undergoing Compton scattering (CS+PE), the photon exits the crystal after undergoing Compton scattering (CS), and the extensions of these for two (2CS+PE \& 2CS), and three or more Compton scatterings ($\geq$3CS+PE \& $\geq$3CS). The number of processes in the three studied configurations per 200 million primary photons of energy 661.7 keV is shown in figure \ref{fig:frequency-cs}. The notation \textit{Deposition status} refers to the entirety of the photon energy being transmitted to the crystal, i.e. the sum of individual energy depositions being equal to the primary photon energy. A criterion of 99.9\% of the original energy is set to counter numerical errors. In the case of $^{137}$Cs, events with the total energy deposition greater or equal to 661.0383~keV are assigned the status \textit{Complete}, whereas other cases are considered \textit{Incomplete}. It should be noted that only \textit{Complete} events should be used in traditional reconstruction methods. 

\begin{figure}[h]
    \centering
    \includegraphics[width=\linewidth]{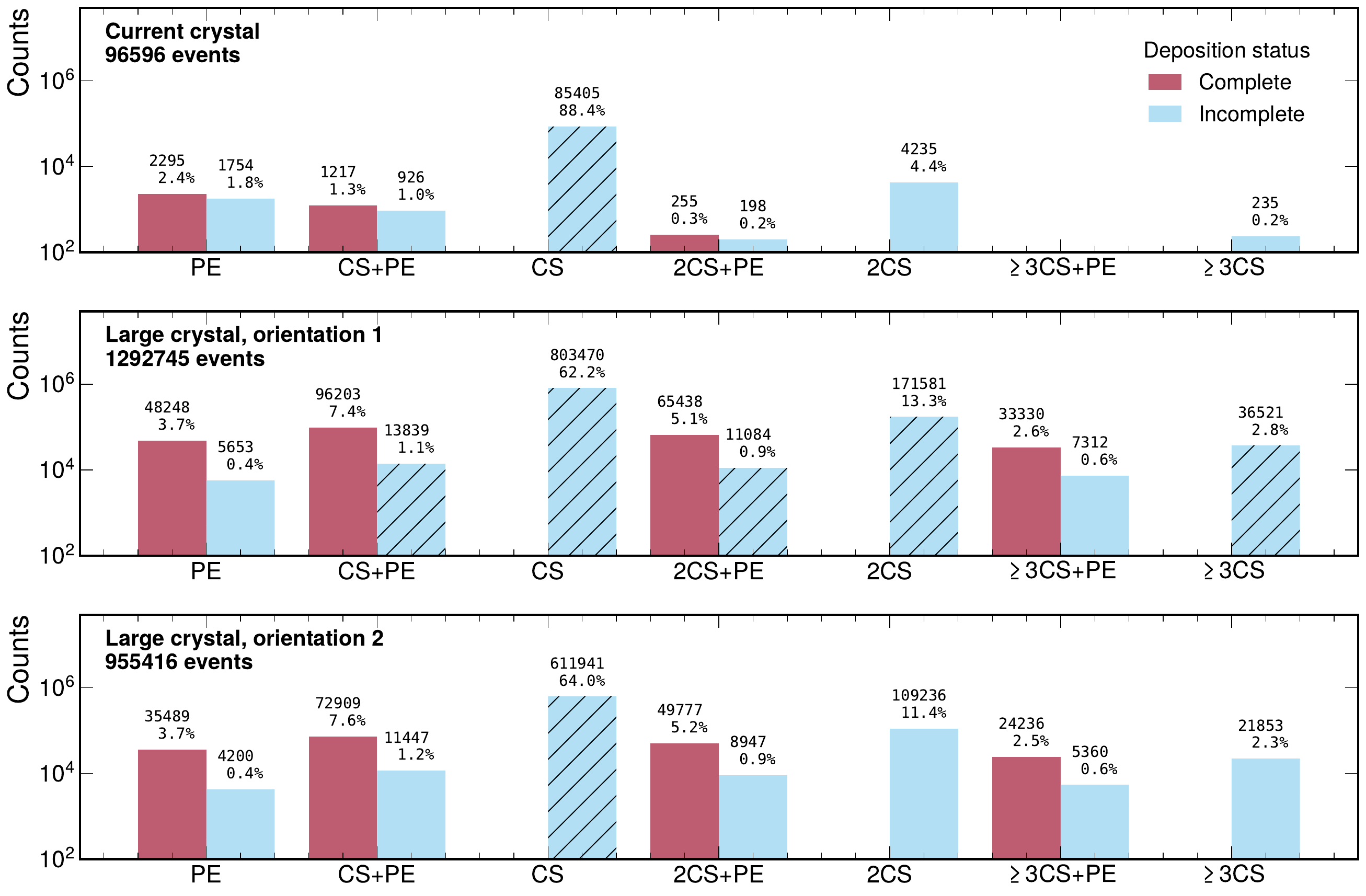}
    \caption{Process frequency in CZT crystals for 1274~keV primary gamma energy. The number and percentage reading above the bar charts show the absolute and relative frequency of the event category.}
    \label{fig:frequency-eu}
\end{figure}

\noindent It is seen that the amount of recorded events is drastically greater for the larger crystals compared to the smaller ones. While configuration 1 provides the highest overall sensitivity, configuration 2 has the largest fraction of golden events at 14.0\%, compared to 13.6\% for configuration 1, and 3.1\% for the small crystals. In other words, the signal purity is the highest for configuration 2. This is thought to be related to events where the secondary photon from a Compton scattering emerges at an angle close to 90 degrees. On the other hand, the absolute number of golden events is the largest for configuration 1, with around 43\% more golden events than configuration 2. This is connected to the higher absorption probability provided by the longer travel distance in the crystal. The process frequency for $^{154}$Eu is shown in figure \ref{fig:frequency-eu}. The relative amount of golden events for $^{154}$Eu is similar to $^{137}$Cs: configuration 1 provides the highest sensitivity, while configuration 2 has slightly higher purity. Interestingly, more events are recorded with $^{154}$Eu than $^{137}$Cs when comparing only configuration 2. This is thought to be caused by the collimator "punch through" phenomenon, where the incoming gamma photons are so energetic, that they can penetrate the thin parts of the collimator with significant efficiency. This effect is detrimental to the reconstruction process.

\section{Compton imaging}
\label{sec:compton}

\begin{figure}[h]
    \centering
    \includegraphics[width=0.49\linewidth]{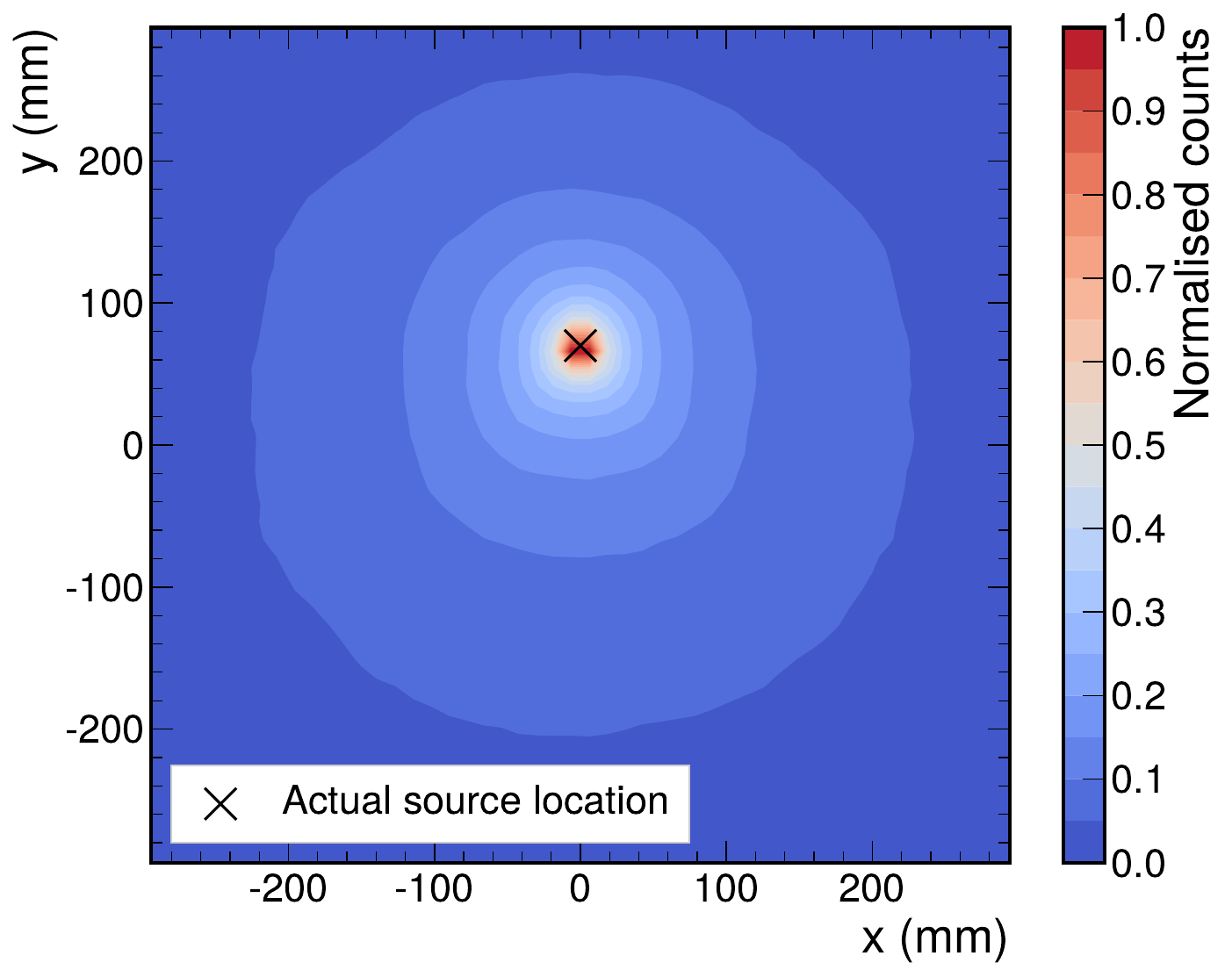}
    \includegraphics[width=0.49\linewidth]{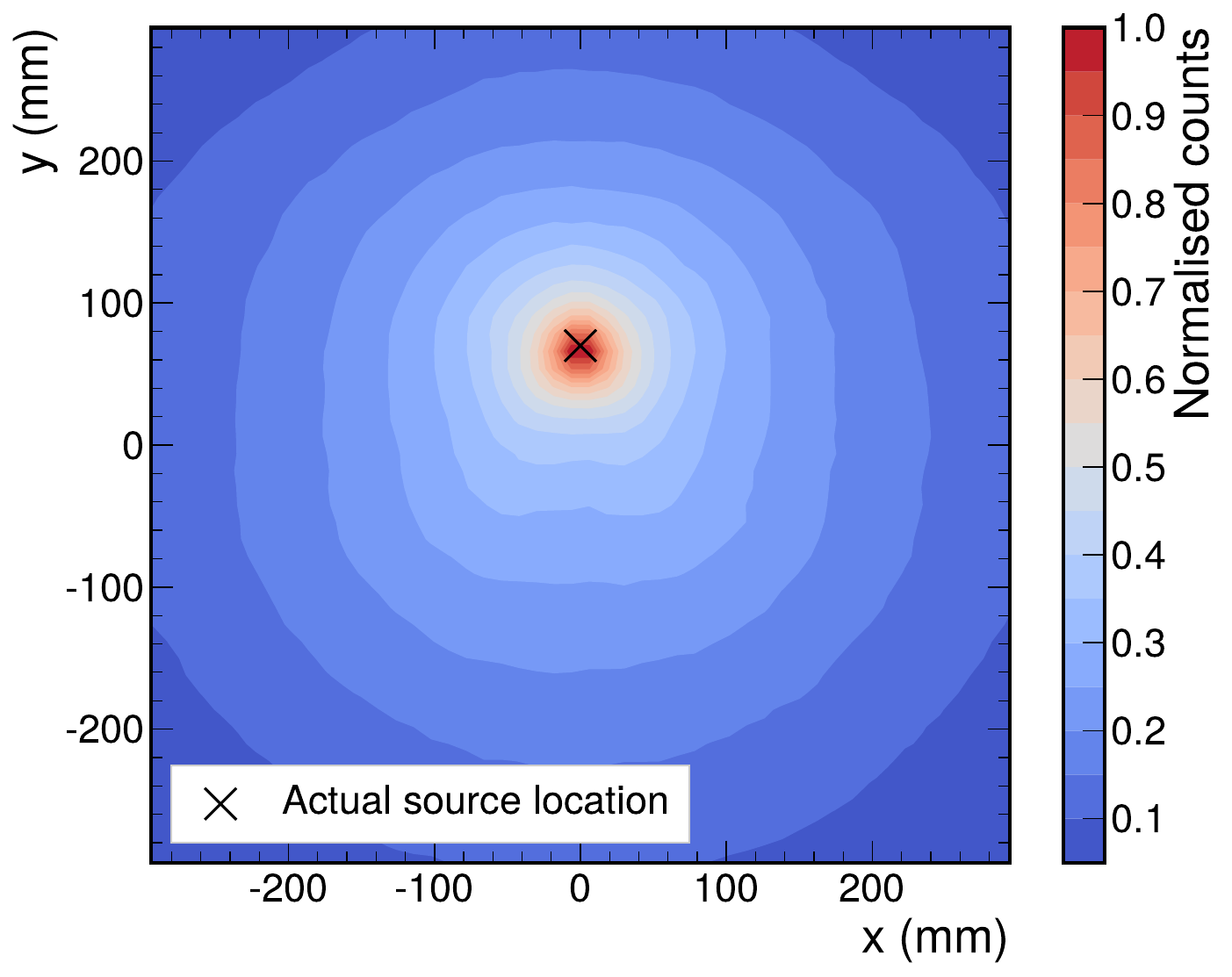}
    \caption{\label{fig:compton}Compton image of a 661.7~keV point source derived using the MC truth (left), and after considering the detector energy and spatial resolution (right). The location of the point source is marked with a "$\mathbf{\times}$" symbol.}
\end{figure}

The collimator design allows imaging 30~cm long axial sections of the fuel assembly at a time. While a more narrow acceptance would potentially improve imaging resolution along the fuel rod, the measurement time would increase. Utilising Compton imaging could enable resolving axial variations in intensity within the imaged area. This option is only available for pixelated devices, which means it is not possible to perform with the current solution. The performance of such methodology is evaluated by simulating a point source at a location $(0, 70~\text{mm}, 0)$ generating photons towards the center of the centermost crystal. The simulation is performed with crystal configuration 1. Two versions of Compton imaging are compared: first assuming an ideal detector with access to the true locations and nature of interactions, and then using realistic values for position and energy resolution. Performance of the "ideal detector" is determined by the statistical nature of the scattering process in CZT, producing a meaningful upper limit to the achievable performance. Events are pre-selected with a condition for the total deposited energy per event. The energy cut is  660~keV for the first case, and 650~keV for the second. In the latter case, the energy and position are smeared using a Gaussian distribution, the energy with 2\% root mean square (RMS) deviation and position with $(\text{pixel pitch})/\sqrt{12}\approx0.635$~mm RMS in the pixelated plane, and 0.357~mm RMS along the depth of the detector based on previous knowledge on the device performance. Since the interaction mechanisms are unknown in a real scenario, events with exactly two "clusters" or separate location of energy depositions are selected. The assumption is that one of these belongs to the Compton scattering and the other one to the photoelectric effect. In both cases, 2 million events passing the criteria are used. This choice is made to have a direct comparison of the quality of the estimation, not being sensitive to the amount of data passing the selection. Figure \ref{fig:compton} shows the direction-of-arrival estimates projected to the z=0 plane according to figure \ref{fig:geometry}.  

\section{Conclusion}
\label{sec:conclusion}

We found the photon detection efficiency of larger CZT crystals to be significantly better than the efficiency of presently installed small crystals. The simulation shows the interaction probability to rise by a factor of 11 for configuration 1, and 7.1 for configuration 2, when measuring the $^{137}$Cs gamma peak at 661.7~keV. For the $^{154}$Eu gamma peak at 1274~keV, the rise is a factor of 13 for configuration 1 and 9.9 for configuration 2. The fraction of "golden events" describing the signal purity relevant for Compton imaging is increased in case of $^{137}$Cs from 3.1\,\% to 13.6\%, and 14.8\% for configurations 1 and 2, respectively.
The similar increase for the $^{154}$Eu gamma peak at 1274~keV is from 1.3\% to 7.4\% and 7.6\%. The overall interaction probability is higher for configuration 1 when measuring $^{137}$Cs, but higher for configuration 2 when measuring $^{154}$Eu possibly due to unwanted effects. Despite configuration 2 having a slightly higher signal purity, the overall efficiency of configuration 1 makes it a more appealing choice.

\acknowledgments

This work was financially supported by NKS (Nordic nuclear safety research) under contract number AFT/NKS-R(24)136/4.

% We suggest to always provide author, title and journal data:
% in short all the informations that clearly identify a document.

\bibliographystyle{unsrtnat}

\end{document}